\documentstyle[11pt]{article}
\newcommand{\be}{\begin{equation}}
\newcommand{\ee}{\end{equation}}
\newcommand{\bea}{\begin{eqnarray}}
\newcommand{\eea}{\end{eqnarray}}
\newcommand{\hA}{\hat{A}}
\newcommand{\hB}{\hat{B}}
\newcommand{\hC}{\hat{C}}
\newcommand{\hD}{\hat{D}}
\newcommand{\hf}{\hat{f}}
\newcommand{\f}{\frac}
\newcommand{\e}{\epsilon}

\newcommand{\g}{\gamma}
\newcommand{\vs}[1]{\vspace{#1 mm}}
\newcommand{\hs}[1]{\hspace{#1 mm}}

\begin{document}

\thispagestyle{empty}

\rightline{hep-th 0110214}

\vs{20}

\centerline{\large\bf On Conifolds and D3-branes}
\vs{10}
\centerline{Ali Kaya\footnote{ali@rainbow.physics.tamu.edu}} 
\vs{5}
\centerline{Center for Theoretical Physics, Texas A\& M University,}
\centerline{College Station, Texas 77843, USA.}
\vs{15}
\begin{abstract}

We search for Ricci flat, K\"{a}hler geometries which are asymptotic to
the cone whose base is the space $T^{11}$ by working out covariantly
constant spinor equations. The metrics we find are singular in the
interior and introducing parallel D3-branes does not form regular event
horizons cloaking the naked singularities. We also work out a
supersymmetric ansatz involving only the metric and the 5-form field
corresponding to D3-branes wrapping over the non-trivial 2-cycle of
$T^{11}$. We find a system of first-order equations and argue that the
solution has an event horizon and the ADM mass per unit volume 
diverges logarithmically.

\end{abstract}
\vs{50}
\pagebreak
\setcounter{page}{1}

The conifold is a 6-dimensional complex manifold  described by a quadric equation in $C^{4}$
\be\label{con}
z_{1}^{2}+z_{2}^{2}+z_{3}^{2}+z_{4}^{2}=0.
\ee
It can be shown that this quadric is a cone whose base is $S^{2}\times
S^{3}$. It is also possible to find a Ricci flat, K\"{a}hler metric on the
conifold \cite{can1}, which may be written as
\be\label{con2}
ds^{2}=dr^{2}+r^{2}ds_{T^{11}}^{2},
\ee
where the Einstein space $T^{11}$ has the topology $S^{2}\times S^{3}$. The Einstein metric of $T^{11}$ can be written explicitly 
\be
ds_{T^{11}}^{2}=\frac{1}{6}\sum_{i=1}^{2}(d\theta_{i}^{2}+\sin 
\theta_{i}^{2}d\phi_{i}^{2})+\frac{1}{9}(d\psi+\sum_{i=1}^{2}\cos\theta_{i}d\phi_{i})^{2}.
\ee
The apex of the cone is singular and there are different ways of removing
the singularity. It is, for instance, possible to $deform$ (\ref{con}) in
such a way that the node is replaced by an $S^{3}$. It is also possible to
rewrite (\ref{con}) by a linear change of variables and then make a
$resolution$, which replaces the node by $S^{2}$. These operations
preserve the Calabi-Yau structure of the conifold \cite{can1}.

\
\

Studying $N$ parallel D3-branes placed at the singularity of the conifold \cite{keh}
one discovers an interesting extension of the $AdS/CFT$ duality
\cite{ads1, ads2, ads3} where the string theory on $AdS_{5}\times T^{11}$
is dual to a certain ${\cal N} = 1$ supersymmetric gauge theory
\cite{wit1, mor1}. The superconformal field theory has the gauge group
$SU(N)\times SU(N)$ and contains chiral superfields with a superpotential.
Introducing $M$ fractional \cite{pol1, doug1} D3-branes, which are indeed
D5-branes wrapped over the collapsed 2-cycle at the singularity
\cite{kl1}, changes the gauge group to $SU(N+M)\times SU(N)$.  This theory
is no longer conformal, and the relative gauge coupling runs
logarithmically \cite{kl1}.

\
\

The supergravity solutions in the presence of fractional D-branes has been
studied in several papers \cite{kl1,ts1,kl2,ts2,ts3, kl3, chr1}. It is
remarkable that, introducing fractional branes changes the geometry in a
controlled way. In the usual D-brane solution the warp factor is the zero
eigenvalue of the Laplacian on the transverse space. Introducing
fractional D-branes, the differential equation picks up a source term and
the the harmonic function is modified so that the warp factor 
becomes \cite{kl3}
\be\label{har}
H= 1 + \frac{Q}{r^{7-p}} + h(r),
\ee
where $p<6$ and 
\be\label{h}
h(r)\sim \begin{cases}{1/r^{10-2p}\hs{5} p = 0,1,2,4; \cr\cr
                       \ln r/r^{4} \hs{8} p=3; \cr\cr
                       \ln r \hs{13} p=5.} \end{cases} 
\ee
The case $p=5$ may be unphysical as discussed in \cite{kl3}.

\
\

For $p<5$, the geometry is asymptotically flat. Removing 
asymptotically flat region by ignoring the constant term in (\ref{har}),  
one can "zoom in" on the low energy dynamics and decouple the interactions 
between the supergravity in the bulk and the gauge theory on the branes. 
For $p=3$, this gives the gravity dual of the $SU(M+N)\times SU(N)$ gauge 
 theory corresponding $M$ fractional and $N$ regular D3-branes \cite{ts1}. In this 
 solution, the 3-form flux is responsible for conformal symmetry breaking 
and indeed the 2-form potential acquires a logarithmic radial dependence 
which implies the logarithmic running of the gauge couplings in the field theory. As $r\to\infty$, the solution is regular and can be used as the gravity dual of $SU(N+M)\times SU(N)$ theory in the UV. However, toward small $r$ one encounters a singularity, which implies that the solution should be modified to describe physics in the IR.

\
\

On the other hand, for $p=3$ it is possible to indicate two difficulties in
obtaining the gravity dual of the gauge theory from the asymptotically
flat solution. The first point is that, due to the special logarithmic
correction to the warp factor in (\ref{h}) the ADM mass per unit volume of the flat
solution diverges logarithmically. Therefore, it is indeed hard
to consider that solution in the space of physical states of the 
supergravity theory. The second difficulty is that, the solution does not
have an event horizon, again due to the special logarithmic correction.
However, as it is well known in the context of $AdS/CFT$ duality, in
taking the scaling limit or ``zooming in'' on the low energy dynamics or
decoupling the asymptotically flat region, the presence of an event
horizon is responsible for the infinite redshift of the energies and plays
the crucial role. Thus, we think that it would be appropriate to consider
the gravity dual of $SU(N+M)\times SU(N)$ gauge theory (the background
with the warp factor (\ref{har}) without the constant term) as the scaling
limit of some other unknown black-brane solution which has a finite mass
and a regular event horizon.

\
\

It is also interesting to consider the fate of the naked conifold
singularity in the presence of D-branes. It is well known that when
parallel D3-branes are placed at the singularity, there forms a regular
event horizon cloaking the singularity. Introducing fractional
D-branes, the story gets complicated, and more fields play a role in the solution. 
Naively, one would still expect
formation of an event horizon. Alternatively, recalling the fact that
the dual gauge theory breaks chiral symmetry in the IR and analyzing the
moduli space, one can replace the singular conifold of the supergravity
background with the deformed conifold from the beginning, and thus both
can solve the singularity problem in the IR and obtain a geometrical
realization of chiral symmetry breaking \cite{kl2}. Finally, it is possible to  
resolve
singularity by adding angular momentum to the supergravity background,
which also reduces the number of supersymmetries \cite{mal1}. 

\
\

Motivated by these recent developments, in this letter we first search for
Ricci flat, K\"{a}hler geometries asymptotic to the cone whose base is the
space $T^{11}$. These spaces can be viewed as the (singular) deformations
or resolutions of the conifold. One may have a purely mathematical
interest in finding such metrics having restricted holonomies. However,
our main concern here is to understand in the context of supergravity
theory how the singularities are modified in the presence of parallel
D-branes. As mentioned above, when the D3-branes are located at the
singularity of the conifold, there forms an event horizon cloaking the
singularity. One may wonder if this is also the case for other singular,
Ricci flat, asymptotically conifold metrics. If it is the case, then one
would hope to take a scaling or near horizon limit of the solution and
obtain gravity duals of certain supersymmetric gauge theories.
Unfortunately, the answer turns out to be negative for the spaces we
consider; in the presence of D3-branes one still encounters either naked
singularities or singular horizons.

\
\

In this paper, we also consider a supersymmetric ansatz involving only the
metric and the 5-form field corresponding to D3-branes wrapping over the
2-cycle of the space $T^{11}$. Recalling that wrapped D$p+2$-branes are
fractional $Dp$-branes, the background can be thought to be related to
fractional D1-branes. Our ansatz differs from the fractional D1-brane
solution of \cite{kl3} where in addition to self dual 5-form field the
dilaton, NS and RR 3-forms acquire non-zero vacuum expectation values. The
2-cycle in our ansatz is a supersymmetric cycle of $T^{11}$ \cite{str1},
and thus one may claim that the D3-branes can wrap it without exciting
other fields.  Existence of a supersymmetric background having only the
metric and the 5-form field supports this claim. Following \cite{ak1, ak2},
we derive a system of first order equations and argue that the ADM mass 
per unit volume diverges logarithmically and the solution has an event 
horizon. 

\
\

Let us consider a 6-dimensional metric of the form
\bea
ds^2&=&f(r)^2dr^2\,+ \,\frac{B(r)^2}{6} \,(d\theta_{1}^{2}+\sin 
\theta_{1}^{2}d\phi_{1}^{2})\,+\, \frac{C(r)^2}{6}\,(d\theta_{2}^{2}+\sin 
\theta_{2}^{2}d\phi_{2}^{2})\nonumber\\
&+& \frac{D(r)^2}{9}(d\psi+{\cal A})^2, \label{conmet}
\eea
where 
\be \label{A}
{\cal A}=\cos\theta_1 d\phi_1 + \cos\theta_2d\phi_2.
\ee
Note that ${\cal A}$ is the one-form potential of the complex
structure\footnote{One may consider a more general potential of the form
${\cal A}=p\cos\theta_1 d\phi_1 + q\cos\theta_2d\phi_2$, where $p$ and
$q$ are integers. However, it turns out that only when $p=q=1$ the metric
admits covariantly constant spinors.} on $S^2\times S^2$. We would like to
determine the unknown functions $f$, $B$, $C$ and $D$ obeying the boundary
conditions
\be
f\to 1, \hs{10} B,C,D\to r\hs{5}\textrm{as}\hs{2}r\to \infty,
\ee
so that the metric (\ref{conmet}) becomes Ricci flat and K\"{a}hler. The boundary conditions make sure that the geometry is asymptotically conic whose base is $T^{11}$. Instead of calculating Ricci tensor, solving second order, coupled differential equations and further imposing a K\"{a}hler structure, we demand existence of a covariantly constant spinor ${\cal \e}$. It is very well known that this implies Ricci flatness and one can also construct a globally well defined and covariantly constant complex structure 
\be\label{complex1}
J_{ab}\,=\, i\,{\cal\e}^\dagger\Gamma_{ab}{\cal\e},
\ee
obeying
\be
J_{a}{}^{b}J_{b}{}^{c}=-\delta_a^b,
\ee
where $a,b,c\,=\,1..6$ are tangent space indices on ({\ref{conmet}). The
last equation can be verified by a Fierz identity. In solving the spinor
equations, we use the $gauge$ covariantly constant
spinors on $S^2\times S^2$ obeying \cite{chr2}
\be\label{cov1}
D_{\alpha}\eta \equiv (\nabla_\alpha+\frac{1}{2}{\cal A}_\alpha)\eta=0
\ee
and 
\be \label{cov2}
J_{\beta\alpha}\Gamma^{\alpha}\eta=i\Gamma_\beta\eta ,
\ee
where the one-form ${\cal A}$ is given in (\ref{A}), 
$(\alpha,\beta)=1,..,4$ are 
tangent space indices and $\nabla_\alpha$ is the covariant derivative on 
$S^2\times S^2$. One can show that ${\cal \e}$ is a covariantly
constant spinor on (\ref{conmet}) provided that it is chosen to be a 
chiral spinor obeying  
\be
{\cal \e}\,=\,e^{-i/2 \psi}\,\eta,
\ee
and 
\bea
\frac{D'}{fD}&=&-\frac{D}{B^2}-\frac{D}{C^2}+\frac{3}{D},\label{eqd}\\
\frac{B'}{fB}&=&\frac{D}{B^2},\label{eqb}\\
\frac{C'}{fC}&=&\frac{D}{C^2}\label{eqc},
\eea
where $'$ denotes derivative with respect to $r$. Note that the chirality
of $\e$ is consistent with (\ref{cov1}) and (\ref{cov2}). One may surprise
by the fact that there are four independent functions and three
differential equations. This is simply a manifestation of the
reparametrization invariance related to the choice of the coordinate $r$
in the metric (\ref{conmet}). One can indeed fix one of the unknown
functions by using this invariance as we will do in a moment.

  \
\

From (\ref{eqb}) and (\ref{eqc}), one finds that
\be\label{g}
B^2=C^2+\g^2,
\ee
where $\g$ is a constant. For $\g=0$ one can fix $r$-reparametrization
invariance by imposing $B=C=r$. Remaining unknown functions $f$ and $D$ can
be solved from (\ref{eqd}) and (\ref{eqb}) to give the following Ricci 
flat metric\footnote{The solution $f=1$ and $D=r$ corresponds to the 
conifold.}
\be
ds^2=\f{r^6}{r^6-1}\,dr^2\,+ \,\f{r^2}{6}\,\, \sum_{i=1}^{2}(d\theta_{i}^{2}+\sin 
\theta_{i}^{2}d\phi_{i}^{2})\,+\,\frac{r^2}{9}
\left(\f{r^6-1}{r^6}\right)(d\psi+{\cal A})^2. \label{met1}
\ee
Another solution can be obtained by letting $r\,\to\,i\,r$ which gives 

\be 
ds^2=\f{r^6}{r^6+1}\,dr^2\,+ \,\f{r^2}{6}
\,\,\sum_{i=1}^{2}(d\theta_{i}^{2}+\sin
\theta_{i}^{2}d\phi_{i}^{2})\,+\,\frac{r^2}{9}\left(\f{r^6+1}{r^6}\right)(d\psi+
{\cal A})^2. \label{met2} 
\ee
Note that (\ref{met1}) and (\ref{met2}) represent two different geometries, i.e. there is no coordinate transformation that will take (\ref{met1}) into (\ref{met2}). 

\
\

For $\g\not=0$, we can parametrize $B$ and $C$ by
\bea
B&=&\g\cosh \rho,\nonumber\\
C&=&\g\sinh \rho. \label{rep}
\eea
Introducing a new radial coordinate $u$ defined by
\be
\f{D}{fdr}=\f{1}{du}, 
\ee
and from (\ref{rep}), (\ref{eqb}) and (\ref{eqd}) we obtain
\bea
D^2&=&\f{e^{6u}}{\cosh^2\rho\sinh^2\rho}\\
e^{6u}&=&\g^2(\sinh^6\rho\,+\,\f{3}{2}\sinh^4\rho\,+\, \textrm{c}')
\eea
where $c'$ is a constant. In terms of $r\equiv\g\sinh\rho$, this gives the 
following Ricci flat metric\footnote{Note that the coordinate $r$ in (\ref{met3}) is 
different from the original radial coordinate.}
 \bea
ds^2&=&K(r)^{-1}\,\,dr^2\,+ \,\frac{(r^2+\g^2)}{6} \,(d\theta_{1}^{2}+\sin 
\theta_{1}^{2}d\phi_{1}^{2})\,+\, \frac{r^2}{6}\,(d\theta_{2}^{2}+\sin 
\theta_{2}^{2}d\phi_{2}^{2})\nonumber\\
&+& K(r)\frac{r^2}{9}(d\psi+{\cal A})^2, \label{met3}
\eea
where
\be
K(r)=\f{(r^6\,\,+\,\,\f{3}{2}\,\g^2\,r^4\,\,+\,\,c)}{r^4\,(r^2\,+\,\g^2)},
\ee
and $c$ is a constant. For $c=0$, (\ref{met3}) becomes the $resolved$ 
conifold metric which is explicitly given in \cite{ts2}. Thus here we 
found that it belongs to a larger 
two parameter family of metrics (\ref{met3}). 

\
\

By construction, the metrics (\ref{met1}), (\ref{met2}) and (\ref{met3})  
are Ricci flat and admit covariantly constant spinors. Thus the holonomy
group of each metric is restricted. Furthermore, one can find a
covariantly constant complex structure by using (\ref{complex1}). In the
tangent space basis $E^r=fdr$, $E^{i_1}=Be^{i_1}$, $E^{i_2}=Ce^{i_2}$ and
$E^D=D(d\psi+{\cal A})$, where $e^{i_1}$ and $e^{i_2}$ refers to tangent
space of $S^2\times S^2$, respectively, the complex structure takes the
standard form
\bea
J_{rD}&=&-J_{Dr}=1,\nonumber\\
J_{i_1i_1'}&=&\e_{i_1i_1'},\nonumber\\
J_{i_2i_2'}&=&\e_{i_2i_2'}.\nonumber
\eea
One can indeed verify that $J_{ab}$ is covariantly constant, and thus  
(\ref{met1}), (\ref{met2}) and (\ref{met3}) are Ricci flat, K\"{a}hler 
metrics.

\
\

Asymptotically, as $r\to \infty$, all three metrics approach to the
conifold (\ref{con2}). All three metrics are also singular in the interior,
except the resolved conifold metric which corresponds to
$c=0$ in (\ref{met3}) and is known to be regular. The metric
(\ref{met1}) is defined for $r\geq1$ and as $r\to1$ $S^2\times S^2$ has a
finite volume but the $U(1)$ bundle parametrized by the coordinate $\psi$
shrinks to zero size forming a singularity. The metrics (\ref{met2}) and
(\ref{met3})  are defined for $r\geq 0$. In (\ref{met2}), as $r\to 0$, the
$U(1)$ bundle expands (therefore the curvatures decrease) but $S^2\times S^2$ 
shrinks to zero size forming a singularity. In (\ref{met3}) and for $c\not =0$, 
although one of the $S^2$'s has a finite volume and the $U(1)$ bundle expands, 
the other $S^2$ factor shrinks to zero size as $r\to 0$ forming a singularity.

\
\

Before introducing D3-branes and studying supergravity solutions, one may be curios about the role played by spheres in the above metrics. Replacing $S^2\times S^2$ with $R^2\times R^2$, one may consider a metric of the form.
\be
ds^2=f(r)^2dr^2\,+ \,B(r)^2\,(dx^idx^i) +\, 
C(r)^2\,(dy^idy^i)\,+\,D(r)^2(d\psi+{\cal A})^2,
\ee
where  
\be 
{\cal A}\,\,=\,\,x^idx^i\,+\,y^idy^i,
\ee
$i=1,2$ and $\psi$ is not necessarily periodic. This metric represents a line
bundle over $R^2\times R^2$.  Working out the covariantly constant spinor
equations one finds the following equations
\bea
\frac{D'}{fD}&=&-\frac{D}{2B^2}-\frac{D}{2C^2},\label{eqd2}\\
\frac{B'}{fB}&=&\frac{D}{2B^2},\label{eqb2}\\
\frac{C'}{fC}&=&\frac{D}{2C^2}\label{eqc2}.
\eea
Compared to (\ref{eqd})-(\ref{eqc}), the only difference (in addition to
the one related to normalization of the metric functions) is that the last
term in (\ref{eqd}) is absent. Nothing that (\ref{eqb2}) and (\ref{eqc2})
imply (\ref{g}), one can again parametrize $B$ and $C$ as in 
(\ref{rep}) and solve the remaining equations to obtain the following
metric
\be\label{eski}
ds^2=4r^4(r^2+\g^2)dr^2\,\,+\,\,
r^2dx^idx^i\,\,+\,\,(r^2+\g^2)dy^idy^i\,\,+\,\,\f{1}{r^2(r^2+\g^2)}(d\psi+{\cal 
A}).
\ee
By construction, this one parameter family of metrics are Ricci flat and
K\"{a}hler. The structure of (\ref{eski}) is very similar to (\ref{met3})
and can be thought to correspond to the limit where the radius of the
spheres blow up. For $\g=0$, (\ref{eski}) has been found in \cite{ak2}, so
here we generalize our previous result.

\
\

Another possible modification is to replace $S^2\times S^2$ with a single
copy of $S^2$ and thus consider a four dimensional geometry.  Starting
with an ansatz of the following form
\be\label{32}
ds^2=f(r)^2dr^2\,+ \,B(r)^2\,(d\theta^2+\sin^2\theta 
d\phi^2)\,+\,D(r)^2(d\psi+p\cos\theta d\phi)^2,
\ee
where $p$ is an integer, and using the Killing spinors of $S^2$, one finds that 
(\ref{32}) admits covariantly constant spinors if
\bea
\frac{D'}{fD}&=&-\frac{pD}{2B^2},\label{eqd3}\\
\frac{B'}{fB}&=&\frac{pD}{2B^2}-\f{1}{B}.\label{eqb3}
\eea
To solve these first order coupled differential equations, we first fix 
$r$-reparametrization invariance by imposing $B=r$. After this 
gauge fixing the non-linear differential equations can be solved exactly which gives the following  metric
\be\label{S}
ds^2=\f{(1+\sqrt{1+r^2})^2}{(1+r^2)}\,\,dr^2\,+ \,r^2\,(d\theta^2+\sin^2\theta 
d\phi^2)\,+\,\f{4r^2}{p^2(1+\sqrt{1+r^2})^2}(d\psi+p\cos\theta d\phi)^2.
\ee 
By construction (\ref{S}) is Ricci flat and K\"{a}hler. The metric is regular except at $r=0$, where there is a conic singularity of the following form
\bea
\textrm{as}\hs{3}r&\to&0\nonumber\\
ds^2&\to&4dr^2+r^2\left(d\theta^2\,+\,\sin^2\theta
d\phi^2\,+\,\f{1}{p^2}(d\psi+p\cos\theta d\phi)^2\right).
\eea
Note that for $p=1$, last three terms combine to form the standard $S^3$ metric given in terms of Euler angles. 

\
\

Now, we would like to introduce 
parallel D3-branes on spaces (\ref{met1}), 
 (\ref{met2}) and (\ref{met3}), where the branes are located at finite 
$r$. We are interested in the fate of the singularities 
in the presence of D3-branes i.e. if there forms event horizons 
possibly cloaking  the singularities. We assume that the metric and the 
self-dual 5-form field of IIB theory have the following form 
\bea
ds^2&=&\hat{A}(r)^2\,ds_4^2\,\,+\,\hat{f}(r)^2dr^2\nonumber\\
 &+&\,\frac{\hB(r)^2}{6} \,(d\theta^{2}+\sin\theta^{2}d\phi^{2})\,
 +\, \frac{\hC(r)^2}{6}\,(d\theta'^{2}+\sin\theta'^{2}d\phi'^{2})
\,+\,\frac{\hat{D}(r)^2}{9}(d\psi+{\cal A})^2, \label{dmet}\\
F&\sim&\,(1+*)\,\,\Omega_2\wedge \Omega_{2'}\wedge (d\psi+{\cal A}),
\eea
where $\Omega_2\wedge\Omega_{2'}$ is the volume form on $S^2\times S^2$
with angular coordinates $(\theta,\phi)$, $(\theta',\phi')$, respectively,
${\cal A}$ is given in (\ref{A}) and $ds_4^2$ is the metric on the flat
4-dimensional world-volume.  This is indeed a natural ansatz to consider,
since the solution corresponding to parallel D3-branes located at the
singularity of the conifold has this form. Note that, $dF=0$ and all but
the Einstein equations are satisfied. To find the unknown functions, we
demand the existence of a Killing spinor on the background which would
then imply Einstein equations as shown in \cite{ak1, ak2}. It is not hard
to see that the Killing spinor equations are satisfied if one chooses the
spinor to be a function of $r$ times the covariantly constant spinor on
6-dimensional transverse K\"{a}hler space and
\bea
\frac{\hD'}{\hf\hD}&=&-\frac{\hD}{\hB^2}-\frac{\hD}{\hC^2}+\frac{3}{\hD}-
\f{q}{\hB^2\hC^2\hD} \label{eqhd}\\
\frac{\hB'}{\hf\hB}&=&\frac{\hD}{\hB^2}-\f{q}{\hB^2\hC^2\hD}, 
\label{eqhb}\\
\frac{\hC'}{\hf\hC}&=&\frac{\hD}{\hC^2}-\f{q}{\hB^2\hC^2\hD}, 
\label{eqhc}\\
\f{\hA}{\hf\hA}&=&\f{q}{\hB^2\hC^2\hD}, \label{eqha}
\eea
where $q$ is proportional to the dyonic charge of the D3-branes. Although 
the coupled differential equations seem to be complicated, 
a simple solution can be found   
\be\label{newsol}
\hA=H^{-1/4},\hs{3}  
\hf=H^{1/4}f,\hs{3}\hB=H^{1/4}B,\hs{3}\hC=H^{1/4}C,\hs{3}\hD=H^{1/4},
\ee
where $f$, $B$, $C$ and $D$ obey (\ref{eqd}), (\ref{eqb}) and 
(\ref{eqc}), and 
\be\label{H}
H'\,=\,-\f{4qf}{DB^2C^2}.
\ee
Therefore, introducing parallel D3-branes the background still preserves
some fraction of supersymmetry of the vacuum, and the geometry is changed
by a warp factor obeying (\ref{H}). It is not very surprising that there
is a solution obeying (\ref{newsol}) and (\ref{H}), since it is well known
that given a Ricci flat 6-dimensional space one can construct the
generalization of the D3-brane solution where the the Ricci flat space
plays the role of the transverse space and the warp factor is a harmonic
function on it. It is easy to see that $H$ is indeed harmonic on 
(\ref{conmet}).

\
\

The solution to (\ref{H}) can be written as 
\be\label{solH}
H=1+\int_{r}^{\infty}\f{4qf}{DB^2C^2}dr,
\ee
 so that as $r\to\infty$, $H\to 1$. Specifically $H=1+O(1/r^4)$ for large
$r$, which shows that the solution has a finite ADM mass per unit volume and
asymptotically the geometry becomes the four dimensional flat world-volume
times the space (\ref{met1}), (\ref{met2}) or (\ref{met3}). One can also show that the background support non-zero D3-brane charge which is conserved and equal to ADM mass per unit volume.

\
\

From (\ref{H}), we see that $H'$ is always negative. (Note that the
functions $f$, $B$, $C$ and $D$ are all positive since they are 
square roots of the metric components.) Therefore, $H$ monotomically
increases as $r$ becomes smaller and smaller. Nothing that $H=1$ at
infinity, an event horizon would finally form if $H$ diverges at some $r$.
However, this does not guarantee the regularity of the event horizon.

\
\

We now consider three metrics (\ref{met1}), (\ref{met2}) and 
(\ref{met3}) separately. From (\ref{met1}), the warp factor can be 
written as 
\be
H(r)=1+\int_{r}^{\infty}\f{4qr}{r^6-1}dr.
\ee
The integral cannot be evaluated in terms of elementary functions, 
but the behavior near $r=1$ can easily be found to be $H\sim-\ln(r-1)$. 
Since $H$ diverges at $r=1$ there forms an event horizon, which turns 
out to be a singular surface. Note that, in (\ref{met1}) there 
was a singularity located at $r=1$, and thus introducing parallel 
D3-branes replaces the naked singularity with a null singular surface. 

\
\

The warp factor corresponding to (\ref{met2}) becomes
\be
H(r)=1+\int_{r}^{\infty}\f{4qr}{r^6+1}dr.
\ee
Contrary to the above case, the integral now converges as $r\to 0$, where 
there is a singularity located in (\ref{met2}).  Therefore, introducing 
D3-branes does not change the presence of the naked singularity in 
(\ref{met2}). 

\
\

The warp factor corresponding to (\ref{met3}) is equal to
\be
H(r)=1+\int_{r}^{\infty}\f{4qr}{r^6+\f{3}{2}\g^2r^4+c}dr.
\ee
As discussed in \cite{ts2}, for $c=0$ the above integral can be evaluated 
exactly. Here we note that as $r\to0$, $H$ diverges as $H\sim1/r^2$, 
therefore there forms an event horizon replacing the naked singularity. 
However, the event horizon turns out to be a singular surface. For 
$c\not=0$, the integral would converge as $r\to0$, thus introducing 
D3-branes does not remove the naked singularity nor it does form an event 
horizon.  

\
\

Till now, we have only considered parallel D3-branes on the spaces
(\ref{met1}), (\ref{met2}) and (\ref{met3}), and determined the fate of
the naked singularities. We found that the presence of D3-branes does not
necessarily imply formation of an event horizon, or if an event horizon
would form it is not necessarily regular. We now would like to consider an
ansatz corresponding to D3-branes wrapping the supersymmetric 2-cycle of
$T^{11}$. Recalling that the wrapped D$p+2$-branes are indeed fractional
D$p$-branes, the ansatz can be thought to be related to fractional
D1-branes. We will comment on this later. For now let us consider an
ansatz of the form,
\bea\label{d3}
ds^2&=&E^2(-dt^2+dx_1^2)+A^2(dx_2^2+dx_3^2)\nonumber\\
&+&\frac{B^2}{6} \,(d\theta^{2}+\sin\theta^{2}d\phi^{2})\,
 +\, \frac{C^2}{6}\,(d\theta'^{2}+\sin\theta'^{2}d\phi'^{2})
+ \frac{D^2}{9}(d\psi+{\cal A})^2, \nonumber\\
F&\sim&\,(1+*)\,\,dx_2\wedge dx_3\wedge(\Omega_2-\Omega_{2'})\wedge  
(d\psi+{\cal A}), \label{5form}
\eea
where $\Omega_2\wedge\Omega_{2'}$ is the volume form on $S^2\times S^2$
with angular coordinates $(\theta,\phi)$, $(\theta',\phi')$, respectively,
and the metric functions $E$, $A$, $B$, $C$ and $D$ depend only on $r$. It
is easy to see that $dF=0$ and all but Einstein equations of IIB theory
are satisfied. The structure of the 5-form field in (\ref{5form})
indicates that the D3-branes wrap over the 2-cycle of $T^{11}$ which is
dual\footnote{The duality between the finite dimensional vector spaces
spanned by the cycles ($C_i$) and the forms ($\omega_i$), which are the
basis of homology and co-homology respectively, is defined with respect to
the cup product $\int_C\omega$.} to $(\Omega_2+\Omega_{2'})$. The
coordinates $t$ and $x_1$ span the remaining two dimensions of the
D3-brane world-volume, which can be thought to correspond (fractional)
D1-branes. The coordinates $x_2$, $x_3$ and $r$ together with the 3-cycle
of $T^{11}$ dual to the three form $(\Omega_2-\Omega_{2'})\wedge
(d\psi+A)$ can be identified as the 6-dimensional transverse space.

\
\

The background has Killing spinors, and thus obey Einstein equations, 
provided
\bea
\frac{D'}{fD}&=&-\frac{D}{B^2}-\frac{D}{C^2}+\frac{3}{D}
-\f{q}{A^2B^2D}+\f{q}{A^2C^2D},\label{frd}\\
\frac{B'}{fB}&=&\frac{D}{B^2}+\f{q}{A^2B^2D}+\f{q}{A^2C^2D},\label{frb}\\
\frac{C'}{fC}&=&\frac{D}{C^2}-\f{q}{A^2B^2D}-\f{q}{A^2C^2D},\label{frc}\\
\f{A'}{fA}&=&-\f{q}{A^2B^2D}+\f{q}{A^2C^2D},\label{fra}\\
\f{E'}{fE}&=&\f{q}{A^2B^2D}-\f{q}{A^2C^2D},\label{fre}
\eea
where $q$ is proportional to the D3-brane charge. We demand that the 
metric functions obey the boundary conditions
\be\label{asy}
f,A,E\to 1, \hs{10} B,C,D\to r\hs{5}\textrm{as}\hs{2}r\to \infty.
\ee
We could not succeed in solving these equations explicitly. In principle,
one can find a perturbative power series solution around flat space, which
would determine the asymptotic behavior of the metric. On the other hand,
the fact that we found a system of first order equations replacing the
second order Einstein equations, would help one to extract some useful
information. Indeed, we will argue that the background has an event 
horizon thus represents black (fractional) D1-branes.

\
\

Linearizing the differential equations around flat space, and fixing
$r$-reprametrization invariance by imposing $f=1$, one finds that the
wrapped D3-branes induce following terms as $r\to\infty$
\be\label{asy2}
A,E=1+q\,\,O(\f{\ln r}{r^4}), \hs{5} B,C=r[1+q\,\,O(\f{\ln 
r}{r^2})],\hs{5}D=r[1+q\,\,O(\f{\ln r}{r^4})].
\ee
Recalling that the two of the transverse directions (corresponding to
coordinates $x_2$ and $x_3$) are smeared in (\ref{d3}) and thus
the real transverse space is four-dimensional, we see from (\ref{asy2})
that the solution supports a logarithmically divergent ADM mass per
unit volume proportional to $q$. As noted in the introduction, a similar
logarithmic divergence is encountered for fractional D3-branes.

\
\

By fixing $r$-reparametrization invariance in a suitable way, one can also 
argue that the solution has an event  horizon. Imposing 
\be\label{fix2}
f=\f{4DB^2C^2}{\hat{r}^5(C^2-B^2)},
\ee
where $\hat{r}$ is a radial coordinate, and from (\ref{fra}), one finds 
that
\be
A^2=1+\f{2q}{\hat{r}^4}.
\ee 
On the other hand (\ref{fra}) and (\ref{fre}) implies $AE=1$, so
\be
E^2=\left(1+\f{2q}{\hat{r}^4}\right)^{-1}.
\ee
Therefore, as $\hat{r}\to0$, $E\to0$, which indicates that there forms an 
event horizon at $\hat{r}=0$. Note that the coordinate $\hat{r}$ is 
different than the coordinate $r$ in (\ref{asy2}). Indeed, one can see 
from (\ref{fix2}) that $f$ fails to approach $1$, as $\hat{r}\to\infty$. 
On the other hand, the fact that as $\hat{r}\to\infty$ $A,E\to1$ indicates 
that $\hat{r}$ is also a suitable radial coordinate such that the 
asymptotic region corresponds to large $\hat{r}$.

\
\

Is this background related to fractional D1-branes? For now, it is
hard to answer this question, since the solution is not known explicitly.  
Recall that fractional D1-branes are D3-branes wrapped over the 2 cycle of
$T^{11}$ $collapsed$ at the conical singularity. Therefore, to argue that
the above background corresponds to fractional D1-branes we need to know
the explicit charge distribution which would give the location of the
wrapped D3-branes. Since the 2-cycle in the solution is supersymmetric,
one may claim that the wrapped D3-branes can be placed at any radial
coordinate. However, when the 2-cycle collapses the curvatures diverge,
the (semi-classical) energy of the wrapped D3-branes vanishes and thus
extra massless modes appear \cite{str1} which indicates that the
supergravity description brakes down. (On the other hand, note that the
the energy of the collapsed D3-branes diverges logarithmically after
integrating out the massless modes \cite{str1}. The fact that the ADM mass
of the gravity background diverges logarithmically indicates that
supergravity still encodes some information about collapsed D3-branes.)
In the case of parallel D3-branes placed at the conifold
singularity, the energy of the D3-branes does not
vanish since they do not wrap over any cycles, and thus effectively they
are point-like objects having no internal excitation or energy on $T^{11}$.  
Thus parallel D3-branes do not give rise to extra massless modes. In
addition, the curvature singularity associated with the conifold is
cloaked by the event horizon justifying supergravity description.

\
\

As mentioned in the introduction, the above background differs from the fractional D1-brane solution of \cite{kl3} where the dilaton, NS and RR 3-form fields acquire non-zero vacuum expectation values. We believe that the solution of \cite{kl3} corresponds to the near horizon limit of the background discussed in this letter.  

\
\

It is very well known that supergravity brakes down when the curvatures
become very large. Therefore, it is difficult to have an appropriate
physical picture of manifolds with naked curvature singularities in the
context of supergravity. One would naturally expect that introducing
parallel D-branes (at or before reaching the singularity)  there would
form an event horizon cloaking the naked singularity.  However, the
examples studied in this paper show that this is not always the case; in
the presence of D-branes one still encounters either naked singularities
or singular horizons. Therefore, the situation is not improved in the
context of supergravity. Of course, conic singularities are important
exceptions to this as in the case of the conifold. However, in general, it
seems supergravity does not offer an appropriate description of D-branes
on curved spaces.

\
\

{\bf Note added:} After the submission of the present work to the e-print archive, we learned the paper \cite{tsytnote} which has some overlap with our paper. We thank A. Tseytlin for pointing this out to us.

\end{document}